\documentclass[letterpaper,english,aps,prb,reprint,superscriptaddress]{revtex4-2}
\usepackage[T1]{fontenc}
\usepackage[latin9]{inputenc}
\usepackage{xcolor}
\usepackage{mathrsfs}
\usepackage{amsmath}
\usepackage{amssymb}
\usepackage{makeidx}
\makeindex
\usepackage{graphicx}
\usepackage{lipsum}

\makeatletter

\usepackage{hyperref}
\hypersetup{
    colorlinks = true,
	allcolors = blue	
}
\usepackage{times}

\makeatother

\usepackage{babel}
\begin{document}
\bibliographystyle{apsrev4-2.bst}
\title{Quarter-quantized thermal Hall effect with parity anomaly }
\author{Yu-Hao Wan}
\affiliation{International Center for Quantum Materials, School of Physics,
Peking University, Beijing 100871, China}
\author{Qing-Feng Sun}
\thanks{Corresponding author: sunqf@pku.edu.cn.}
\affiliation{International Center for Quantum Materials, School of Physics,
Peking University, Beijing 100871, China}
\affiliation{Hefei National Laboratory, Hefei 230088, China}

\begin{abstract}
We show that in the proximity of s-wave superconductors, the magnetic topological surface states can transform into Majorana surface state, featuring a single gapless Majorana cone with parity anomaly when the superconducting pairing gap matches the surface magnetization gap.
The emergence of $N=1/2$ Majorana chiral edge current is observed at the boundaries between the gap region and the gapless region.
Additionally, in systems with a single gapless Majorana cone,
a quarter-quantized thermal Hall conductance appears under the dephasing. By mapping the system to a conductor-network model, we identify the appearance of 1/4 chiral heat channels as the cause of the quarter-quantized Hall thermal conductance.
We observe the stability of this quarter-quantized thermal conductance under temperature variations, serving as a distinctive feature indicating the presence of a single gapless Majorana cone in the system.
Our models can be experimentally realized using magnetic topological insulators or iron-based superconductors.
\end{abstract}
\maketitle

\section{\label{sec:level1}Introduction}
Finding a single gapless Dirac
cone of fermions has been a persistent problem in condensed matter
physics \citep{murakami2007phasetransition,armitage2018weyland,yang2014classification}.
Massless Dirac fermions possess parity symmetry, and the coupling with
gauge fields introduces an infinitesimally small mass term to break
parity symmetry through a regularization process, which results in
the emergence of a Chern-Simons theory, known as parity anomaly \citep{redlich1984gaugenoninvariance,redlich1984parityviolation,semenoff1984condensedmatter}.
In recent years, various systems with a single gapless Dirac cone
have been proposed to investigated the parity anomaly \citep{bottcher2019survival,chang2013experimental,zhang2017anomalous,mciver2020lightinduced,lu2018topology,ozawa2019topological,jotzu2014experimental,zhou2022transport,xu2014observation,tokura2019magnetic,ning2023robustness}.
For example, Haldane model achieving a single massless Dirac fermion
by finely tune the band gap of one valley to close in a honeycomb
lattice, while keeping another valley open \citep{haldane1988modelfor}.
Another remarkable approach involves semi-magnetic topological insulators.
In this system, a gapped Dirac cone emerges on one surface due to
the broken of the local time reversal symmetry, while the Dirac cone
on the opposite surface remains gapless \citep{tokura2019magnetic}.
In systems with a single gapless Dirac cone, the parity anomaly leads
to a half-integer quantized Hall conductance, as predicted by the
anomaly-induced Chern-Simons theory \citep{haldane1988modelfor,semenoff1984condensedmatter,lapa2019parityanomaly,burkov2019diracfermion}.
Recently, experimental observations have confirmed the existence
of half-integer Hall conductance in semi-magnetic topological insulators
 \citep{mogi2022experimental}. To describe systems where both massive
and massless Dirac fermions coexist, a concept known as the ``parity
anomaly semimetal" has been proposed \citep{fu2022quantum}.
However, all previous models have been based on electronic systems,
and the exploration of how to realize parity anomaly in Majorana systems
has yet to be investigated.

Majorana fermion can be interpreted as the splitting of a complex
fermionic field of an electron into real and imaginary parts \citep{sato2016majorana},
thus massless Majorana fermion also satisfy the Dirac equation. While
Majorana systems cannot couple with electromagnetic gauge fields,
they can, however, couple with gravity fields \citep{furusaki2013electromagnetic}.
When a massless Majorana fermion couple a gravity field, parity anomaly
results in a gravitational Chern-Simons term \citep{furusaki2013electromagnetic,sekine2021axionelectrodynamics,sato2016majorana,wang2011topological}.
This leads to a quarter-quantized thermal Hall conductance, expressed
as $\ensuremath{{\kappa_{{\rm {H}}}}={\mathop{{\rm sgn}}}(M)\frac{1}{4}\frac{{\pi^{2}k_{{\rm {B}}}^{2}}}{{3h}}\mathcal{T}_0 }
={\mathop{{\rm sgn}}}(M)\frac{1}{4} \ensuremath{{\kappa_{{\rm {0}}}}}
$,
where $\ensuremath{{\kappa_{{\rm {0}}}} =\frac{{\pi^{2}k_{{\rm {B}}}^{2}}}{{3h}}\mathcal{T}_0 }$ is the quantum thermal conductance,
$\mathcal{T}_0$ is the temperature and
$M$ is the mass for the regulator \citep{furusaki2013electromagnetic,sekine2021axionelectrodynamics}.
To achieve parity anomaly in a Majorana system, a system with a single
gapless Majorana cone is required. In time-reversal-invariant (TRI)
topological superconductors (TSC), the surface hosts a singular Majorana
cone \citep{qi2009timereversalinvariant,sato2017topological,sato2016majorana}.
A naive analogy is drawn to the case of semi-magnetic 3D topological
insulators (TIs) where breaking time reversal on one surface achieves parity
anomaly in a Majorana system. However, the absence of bulk materials
confirmed as TRI TSC poses experimental challenges for all TRI TSC-based
approaches \citep{sato2016majorana}.

In this paper, we propose an approach to realize parity anomaly in
Majorana system without the need for TRI TSC bulk materials.
By introducing superconducting pairing through proximity effects on a magnetic topological
surface state, a massless Majorana fermion will emerge when the superconducting
pairing gap matches the surface magnetization gap. Additionally, by
constructing a system featuring only a single Majorana cone, we show
that an $N=1/2$ Majorana chiral edge current emerges on the boundary
between massless Majorana fermion and massive Majorana fermion. Utilizing
non-equilibrium Green's function (NEGF) calculations for a six-terminal
Hall bar system, we identity that the parity anomaly leads to a quarter-Hall
thermal conductance plateau under the influence of dephasing. Finally,
by mapping the system to a conductor-network model, we demonstrate
the connection between the appearance of the 1/4 Hall thermal conductance
plateau and the presence of quarter-quantized chiral
Majorana channels.

The paper is structured as follows.
In Sec. \ref{sec:level2}, by using a low-energy effective model,
we provide a physical depiction of achieving a single gapless Majorana cone.
Sec. \ref{sec:level3} employs numerical methods to illustrate the existence of
$N = 1/2$ Majorana edge chiral current in the Majorana parity anomaly system.
In Sec. \ref{sec:level4}, we calculate the 1/4 Hall thermal conductance in the system under dephasing using NEGF. Besides, by solving a conductor-network model, we establish the relationship between 1/4 Hall thermal conductance and the appearance of 1/4 quantized chiral channels. The stability of the Hall thermal conductance under temperature variations is investigated in Sec. \ref{sec:temperature}.
We address experimental implementation and conclude with a summary in Sec. \ref{sec:discuss}.
Additional computational details and supplementary figures are provided in Appendices \ref{network} to \ref{mu}.

\begin{figure}
\begin{centering}
\includegraphics[scale=0.3]{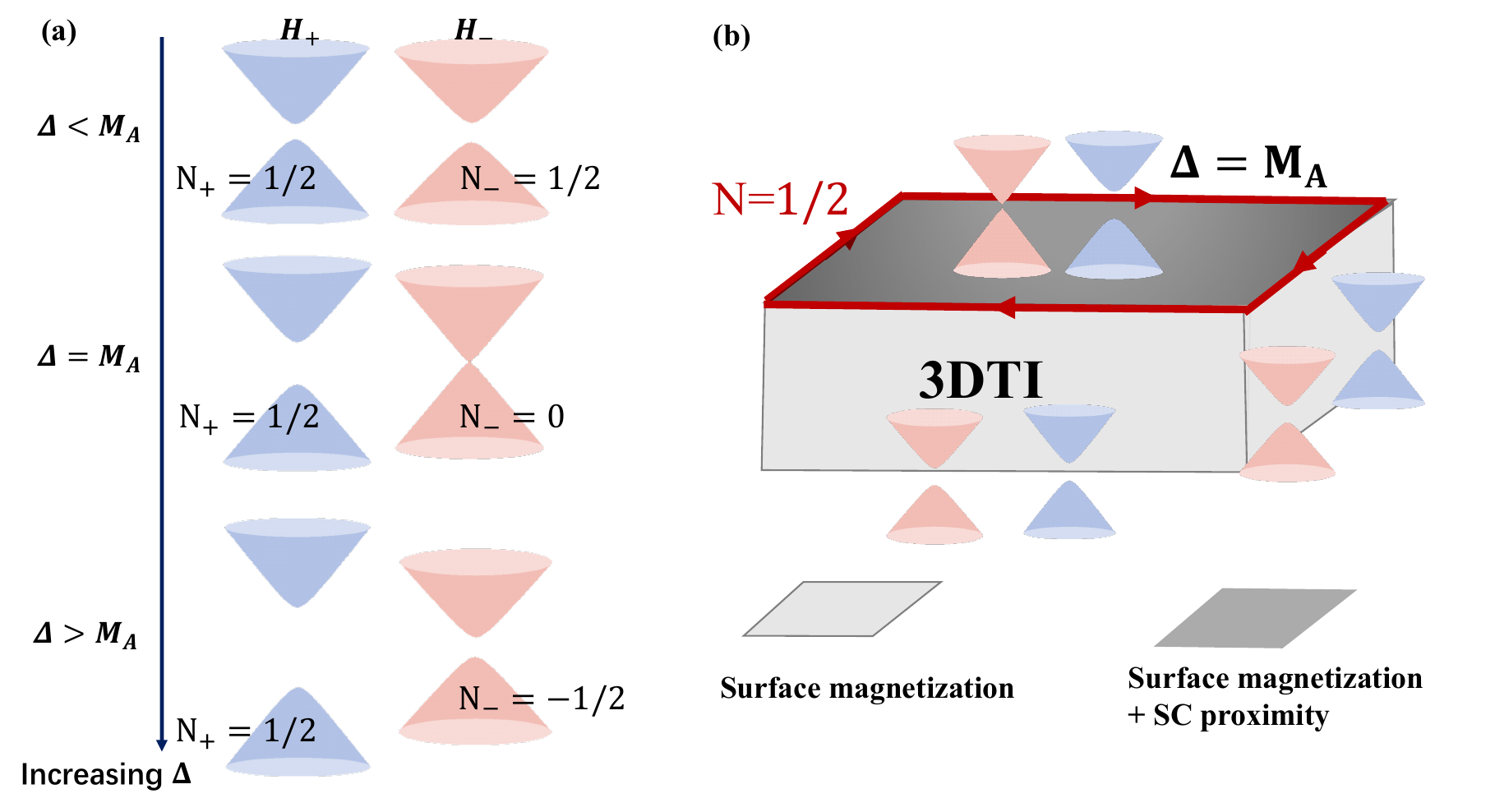}
\par\end{centering}
\caption{\label{fig:1}(a) Schematic diagram of the changes
in the two Majorana cones on the magnetic topological surface
as the superconducting pairing potential $\Delta$ increases,
with the Majorana cones corresponding to $H_{+}$ and $H_{-}$
represented in blue and red, respectively. (b) Schematic diagram of
the system with single gapless Majorana cone on the top surface, with red
and blue cones representing Majorana cones corresponding to $H_{+}$
and $H_{-}$, respectively.}
\end{figure}

\section{\label{sec:level2}Single massless Majorana fermion and parity anomaly}

First, starting from the low-energy model of
the surface of the 3D TI,
we present a physical picture for realizing single gapless Majorana cone.
The surface of the 3D TI hosts a single gapless Dirac cone, with low-energy effective
Hamiltonian expressed as $H_{surf}(k)=A(k_{x}s_{x}+k_{y}s_{y})$
on the basis ($c_{\mathbf{k}\uparrow}, c_{\mathbf{k}\downarrow}$),
where $\ensuremath{{s_{i=x,y,z}}}$ denotes Pauli matrices on the spin space
and $A$ is a parameter related to the Fermi velocity. A gapless Dirac
cone can be interpreted as two gapless Majorana cones in the Bogoliubov-de
Gennes representation \citep{chung2009detecting,qi2011topological}.
However, the presence of two Majorana cones with different chiralities
on the same surface contributes anomaly terms with opposite signs,
resulting in an overall absence of anomaly. Achieving parity anomaly
of Majorana system based on 2D topological surface states would be
possible if there exists a method to selectively open the gap of one
Majorana cone while keeping the other gap closed. Introducing both
magnetization and superconductivity, the surface Hamiltonian in the
Majorana basis $\ensuremath{{\Psi_{k}}}$ is block-diagonalized into
two components: ${{ {H}}_{\pm}}{ {=A}}\left({{{\rm {k}}_{{\rm {x}}}}{s_{{\rm {x}}}}\pm{{\rm {k}}_{{\rm {y}}}}{s_{{\rm {y}}}}}\right){\rm {+}}\left({\Delta\pm{M_{A}}}\right){s_{\rm{z}}}$.
Here, $\ensuremath{\Delta}$ represents the superconducting pairing
potential, $\ensuremath{{M_{A}}}$ signifies the surface magnetization.
The Majorana basis $\ensuremath{{\Psi_{k}}}$ is defined as: $\Psi_{\mathbf{k}}=\ensuremath{\frac{1}{\sqrt{2}}\left(c_{\mathbf{k}\uparrow}+c_{-\mathbf{k}\downarrow}^{\dagger},c_{\mathbf{k}\downarrow}+c_{-\mathbf{k}\uparrow}^{\dagger},-c_{\mathbf{k}\downarrow}+c_{-\mathbf{k}\uparrow}^{\dagger},-c_{\mathbf{k}\uparrow}+c_{-\mathbf{k}\downarrow}^{\dagger}\right)^{T}}$,
where $\ensuremath{c_{{\bf {k}}\uparrow/\downarrow}^{\dagger}}$ and
${c_{{\bf {k}}\uparrow/\downarrow}}$ are the creation and annihilation
operators for electrons \citep{yan2021amajorana,qi2011topological,qi2010chiraltopological}.

In this scenario, $\ensuremath{{H_{\pm}}}$ can be interpreted as
2D massive Majorana fermions with different chirality on the surface,
characterized by masses $\ensuremath{\Delta\pm{M_{A}}}$.
This model can be considered analogous to the Haldane model, which features massive
Dirac fermions of opposite chirality in the K and K' valleys of graphene \citep{haldane1988modelfor}.
The mass term induces the opening of gaps in the Majorana cone $\ensuremath{{H_{\pm}}}$
(as illustrated in the Fig. \ref{fig:1}(a)), with the corresponding
gap's Chern number given by $N_{\pm}=\frac{1}{2}{\mathop{{\rm sgn}}}\left({M_{A}\pm\Delta}\right)$ \citep{yan2021amajorana}.
For convenience in the following discussion, we consider $M_{A}>0$ and $\Delta>0$.
When $\Delta<M_{A}$, both gaps in $\ensuremath{{H_{\pm}}}$ have Chern numbers of 1/2,
resulting in a total Chern number $\ensuremath{N={N_{+}}+{N_{-}}=1}$.
As $\text{\ensuremath{\Delta}}$ increases beyond $M_{A}$, accompanied
by the gap closing and reopening of the Majorana cone corresponding
to $H_{-}$, the Chern number in the corresponding gap changes from
$1/2$ to $-1/2$, yielding a total Chern number $N=0$.
When $\Delta=M_{A}$, the Majorana cone corresponding to $H_{-}$ becomes precisely gapless,
giving rise to the coexistence of massless 2D Majorana fermions with
massive 2D Majorana fermions on one surface, resulting in a total
Chern number of 1/2. Introducing an out-of-plane magnetization $M_{A}$
on all surfaces of a 3D TI and a superconducting pairing potential
$\Delta$ specifically on the top surface, where $\Delta=M_{A}$,
results in a system featuring a single gapless Majorana cone
(see Fig. \ref{fig:1}(b)). Drawing parallels with a system possessing
a single gapless Dirac cone \citep{zou2022halfquantized,gong2023halfquantized,fu2022quantum},
a Majorana edge chiral current with $N=1/2$ emerges on the boundary
of the top surface due to the parity anomaly.

\begin{figure}
\begin{centering}
\includegraphics[scale=0.33]{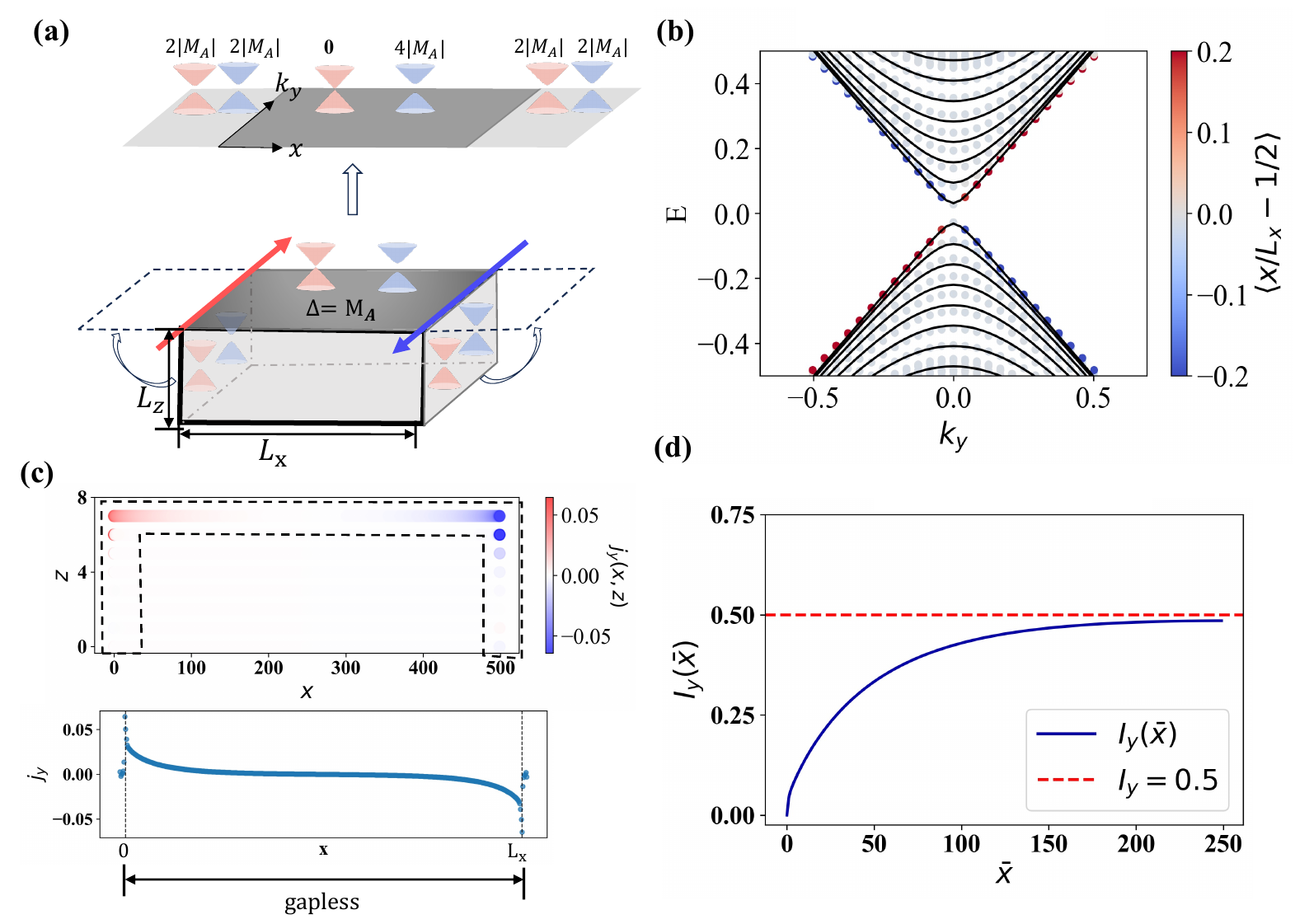}
\par\end{centering}
\caption{\label{fig:2}(a) Schematic diagrams for the 3D TI (lower)
and the unfolding of its surface into a quantum well (upper),
with red and blue cones representing Majorana cones
corresponding to $H_{+}$ and $H_{-}$, respectively.
(b) The analytical results (black lines) and numerical
results (scatter plot) of the model's band structure, with colors
indicating the average displacement relative to the center for each
Bloch state $\ensuremath{\left\langle {x/L_{x}-1/2}\right\rangle }$.
We set $L_{x}=50$, $L_{z}=5$ in the calculation.
(c) The upper panel is the distribution of Majorana current density in the x-z plane.
The lower panel describes the changes in surface current corresponding to the dashed black box in the upper panel.
(d) The variation in Majorana current flux as the summation range $\overline{x}$
increases. {  In both (c) and (d), we have set $L_{x}=500$, $L_{z}=8$ and $\eta  = 0.008$.}}
\end{figure}

\section{\label{sec:level3}$\textit{N=1/2 }$Majorana edge chiral current}
Next,
we use numerical methods to demonstrate the existence of the $N=1/2$
Majorana edge chiral current. The Hamiltonian of a 3D TI in a cubic
lattice is given by \cite{wan2024magnetizationinduced}
\begin{equation}
\mathcal{H}_{\mathrm{TI}}=(m-6B)s_{0}\sigma_{z}+\sum_{i}(2Bs_{0}\sigma_{z}\cos k_{i}+As_{i}\sigma_{x}\sin k_{i}),\label{eq:1}
\end{equation}
where $m$, $B$, and $A$ are the model's parameters, and $k_{i}$
is the momentum with $i=x,y,z$.
$s_{i}$ and $\sigma_{i}$ are the
Pauli matrices on the spin and orbital spaces. By choosing parameters
$m=1$ and $B=0.6$, the 3D TI is placed in a nontrivial topological
phase. Using Eq.(\ref{eq:1}), we establish a lattice model with open
boundary conditions in the x and z directions while maintaining translational
symmetry in the y direction. The widths in the x and z directions
are denoted as $L_{x}$ and $L_{z}$, respectively (see Fig. \ref{fig:2}(a)).
Then, we introduce an out-of-plane magnetization $M_{A}$ on all surfaces.
Additionally, on the top surface, a superconducting pairing potential $\Delta$
arises through proximity to an s-wave superconductor. With $\Delta=M_{A}$,
a gapless Majorana cone (shown in red) emerges on the top surface,
along with a gapped Majorana cone featuring a gap of $4M_{A}$ (blue
cone on the top surface of Fig. \ref{fig:2}(a)).
The remaining surfaces exhibit two Majorana
cones with gaps of $2M_{A}$ each. At the boundary of the top surface,
a Majorana chiral edge current with ${N}=1/2$ (depicted by
red and blue arrows in Fig. \ref{fig:2}(a)) appears,
and we will delve into this phenomenon in the subsequent discussion.

The surface of this model can be effectively unfolded into a quantum
well model (upper subplot in Fig .\ref{fig:2}(a)).
Due to the absence of coupling between the Majorana cones corresponding to $H_+$ (blue cone) and $H_-$ (red cone), they can be treated independently. Since the Majorana cone associated with $H_+$ is gapped on all surfaces and exhibits no states in the energy window of interest (near E=0), our focus can be directed towards the quantum well model formed by the red cone, with massive Majorana cones at both ends and a gapless
Majorana cone in the middle.
To analyze the energy spectrum of this model, we can obtain analytical expressions by considering energy scales much smaller than the gap of the massive cones:
$\ensuremath{{\varepsilon_{n}}\left({k_{y}}\right)=A\sqrt{k_{y}^{2}+{{[\frac{{\pi\left({n+1/2}\right)}}{L_{x}}]}^{2}}}\left({n\in\mathbb{N}}\right)}$ \citep{zou2022halfquantized}.
The finite-size effect leads to an energy gap inversely proportional
to the width $L_{x}$. Fig. \ref{fig:2}(b) demonstrates good agreement
between the numerical and analytical energy spectra, particularly
in the low-energy region. We compute the average displacement relative
to the center for each Bloch state: $\ensuremath{\left\langle {x/L_{x}-1/2}\right\rangle }$,
marked with colors in Fig. \ref{fig:2}(b). States on the left (red)
and right (blue) exhibit opposite velocities, indicating the chiral
nature of the system. { Furthermore, we calculate the distribution of Majorana current density at energy $E$ in the $x-z$ plane}
with $\ensuremath{r=\left({x,z}\right)}$:
\[
\ensuremath{\begin{aligned} & J_{y}(E,\boldsymbol{r})\\
 & =-\frac{e}{\pi h}\int_{-\pi}^{\pi}\operatorname{Im}\operatorname{Tr}\left[\frac{\partial H\left(k_{y}\right)}{\partial k_{y}}G_{k_{y}}^{R}(E,\boldsymbol{r},\boldsymbol{r})\right]dk_{y},
\end{aligned}
}
\]
{ where $G_{k_{y}}^{R}\left({E,{\bf {r}},{\bf {r\prime}}}\right)$ is
the retarded Green's function for $H(k_{y})$, which can be written as $G_{{k_y}}^R\left( {E,{\bf{r}},{\bf{r}}\prime } \right) = \left\langle {\bf{r}} \right|{\left( {E - H\left( {{k_y}} \right) + i\eta } \right)^{ - 1}}\left| {{\bf{r\prime}}} \right\rangle$.}
 The upper panel in Fig. \ref{fig:2}(c) shows 
the distribution of $J_{y}(E=0)$ on the $x-z$ interface.
The opposite-directional current density exists at the left and right
boundaries of the top surface, further confirming the chiral nature
of the system. The current density exhibits distinct decay patterns
in the gapless top surface region and the gap left and right regions,
to better observe the decay trend of current density at the boundaries of the
two regions (gap and gapless), see the lower panel in Fig. \ref{fig:2}(c). 
The current density rapidly decays in the gap region, while it decreases
slowly in the gapless region. As shown in Fig. \ref{fig:2}(d), the
Majorana current flux $\ensuremath{{I_{y}}\left({\bar{x}}\right)=\int_{0}^{{L_{z}}}{dz}\,\,\int_{0}^{\bar{x}}{dx}{J_{y}}\left({x,z}\right)}(\bar{x}<L_{x}/2)$
gradually converges to $1/2$ with increasing $\overline{x}$, { demonstrating
the presence of the $N=1/2$ Majorana current.} { Furthermore, since a chiral Majorana fermion corresponds to half of a complex fermion and contributes half-quantized thermal Hall conductance, we can establish the relationship between Hall thermal conductance and Chern number $N$ as: $\kappa_{xy}=\frac{N}{2}\kappa_0$ \cite{yang2022halfinteger,yan2021amajorana,chung2009detecting}.
Considering the Majorana edge current with $N=1/2$ here, we can expect the exhibiting of a quarter-quantized Hall thermal conductance along with a quarter-quantized chiral heat channel. This will be demonstrated in Section \ref{sec:level4}.}

\begin{figure}
\begin{centering}
\includegraphics[scale=0.35]{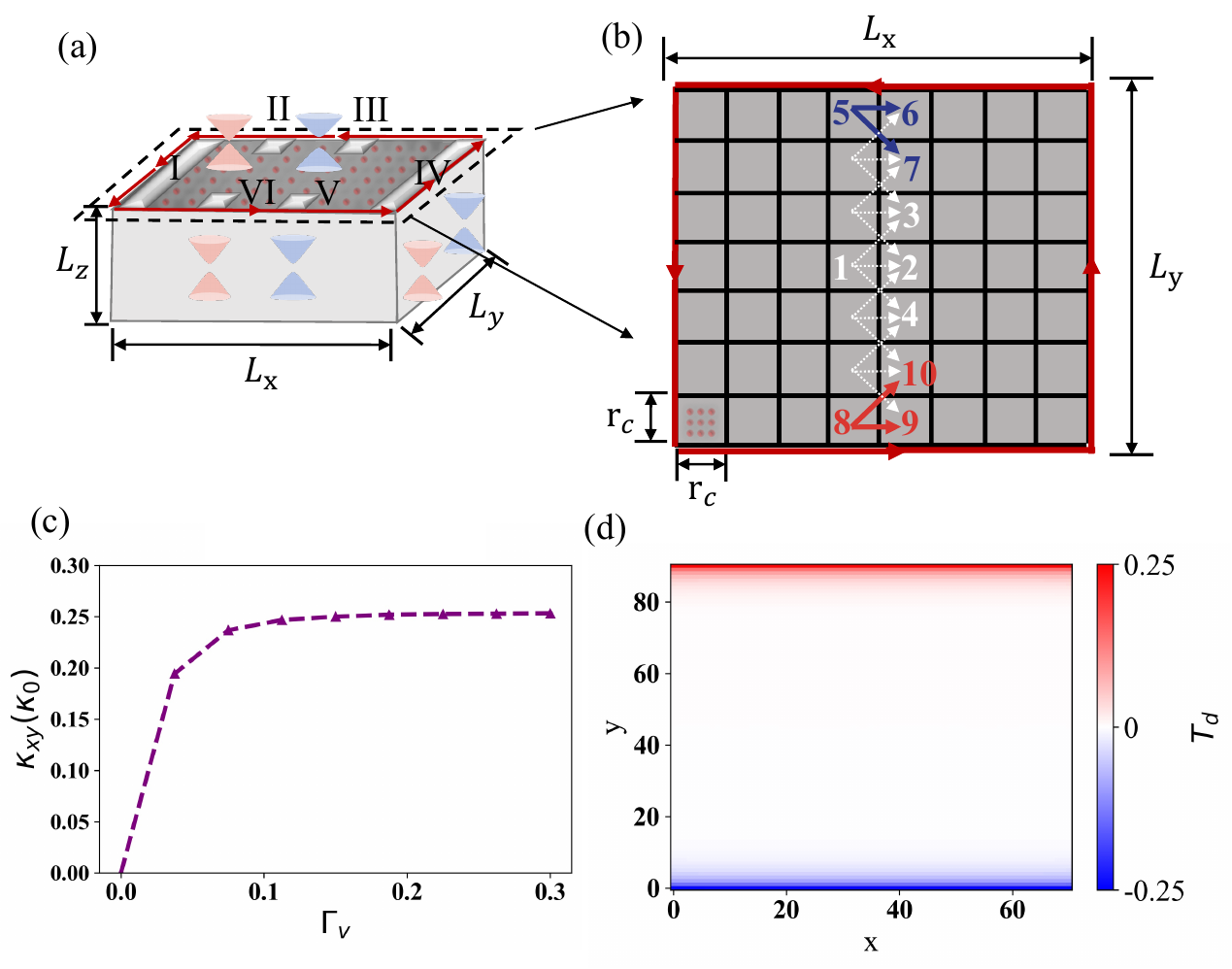}
\par\end{centering}
\caption{\label{fig:3} (a) Illustration of a 3D TI with open boundary conditions
in the x, y, and z-direction, featuring ports attached to the top surface
(highlighted in metallic shade). The red and blue cones correspond
to the Majorana cones on the surface. Red balls here represent the
virtual leads. (b) The red balls on the top surface in (a) are divided
into black boxes, with each black box containing $r_{c}\times r_{c}$
red balls. (c) Variation of thermal Hall conductance with the strength
of the dephasing strength $\Gamma_{v}$. (d) The spatial distribution of the difference in transmission
coefficients $T_{d}$ with $\Gamma_{v}=0.2$.
Here, we set $M_{A}=0.5$ and $\Delta=0.5$.}
\end{figure}

\section{\label{sec:level4}Quarter-quantized thermal Hall conductance}

Thermal transport is very useful for studying Majorana systems
because it effectively avoids the situation where
superconducting systems cannot couple with electric field \citep{furusaki2013electromagnetic,wang2011topological,yang2022halfinteger,tang2019quantized}.
In a 2D chiral topological system, the presence of 1D Majorana fermions
at the edge contributes a half-integer thermal Hall effect to the
system. Recently, the half-integer thermal Hall effect has been observed
in the fractional quantum Hall system at $v=5/2$ \citep{banerjee2018observation}and
Kitaev material $\text{\ensuremath{\alpha-RuCl_{3}}}$ \citep{kasahara2018majorana}. 
Subsequently, we will illustrate the emergence of a quart-quantized
thermal Hall conductance in our system as a consequence of parity
anomaly.

Utilize the cubic lattice model introduced in Sec. \ref{sec:level3}, employing open
boundary conditions in the xyz-direction, where $L_{x}=L_{y}=110$
and $L_{z}=5$. With the superconducting pairing potential $\text{\ensuremath{\Delta}}$
on the top surface equal to the surface magnetization $M_{A}$, a
single gapless Majorana cone emerges on the top surface, results in
a 1/4 Hall thermal conductance in the system due to the parity anomaly.
To investigate surface transport, we introduce a six-terminal Hall-bar
device (see Fig.\ref{fig:3}(a)), with ports I and IV designated as
thermal flow ports, and ports II, III, V, and VI as temperature ports.
Dephasing effect is simulated by introducing B\"{u}ttiker virtual
leads at each lattice point on the top surface (represented by red
spheres in Fig. \ref{fig:3}(b)) \citep{buttiker1986fourterminal,xing2008influence,buttiker1988symmetry}.
Utilizing the Landauer-B\"{u}ttiker formula, the heat current flowing
in the lead-n at low temperatures can be expressed as: \citep{long2011quantum,lambert1993multiprobe,yan2021amajorana}
\begin{equation}
\ensuremath{Q_{n}=\sum_{m\text{\ensuremath{\neq n}}}\left(T_{mn}\mathcal{T}_{n}-T_{nm}\mathcal{T}_{m}\right)
\ensuremath{{\kappa_{{\rm {0}}}}}},\label{eq:2}
\end{equation}
where $\mathscr{\mathcal{T}}_{m}$ is the temperature in lead $m$,
and $T_{nm}^ {}\left({E=0}\right)=T_{nm}^{ee}\left({E=0}\right)+T_{nm}^{CAR}\left({E=0}\right)$.
Here $T_{nm}^{ee\text{/CAR}}\left({E=0}\right)$ denotes the transmission
coefficient of electrons and the cross Andreev process with energy
$E$ from Lead-$m$ to Lead-$n$ respectively. All these transport
coefficients are calculated as:\cite{yan2019electrically} $T_{nm}(E)=\operatorname{Tr}\left[\Gamma_{n}^{e}\mathbf{G}^{R}\Gamma_{m}^{e}\mathbf{G}^{A}\right]$,
$T_{nm}^{CAR}(E)=\operatorname{Tr}\left[\Gamma_{n}^{e}\mathbf{G}^{R}\Gamma_{m}^{h}\mathbf{G}^{A}\right]$,
where $\Gamma_{n}$ is the line-width function of the $n$-th port,
which, in the wide-band approximation, is a constant, and $\mathbf{G}^{R}(E)=\left[\mathbf{G}^{A}\right]^{\dagger}=\left[(E+\mathbf{i}\eta)\mathbf{I}-\mathbf{H}-\sum_{n}\Sigma_{n}^{R}\right]^{-1}$.
The self-energy term is $\Sigma_{n}^{R}=-\frac{\mathbf{i}}{2}\Gamma_{n}$.
For virtual leads, the self-energy $\Sigma_{n}^{R}=-\frac{\mathbf{i}}{2}\Gamma_{v}$,
where $\Gamma_{v}$ is the decoherence strength \citep{xing2008influence}.

When there is a heat current $Q_{x}$ from lead I to lead IV, by solving
Eq.(\ref{eq:2}), the Hall thermal resistance and longitudinal thermal
resistance are obtained as $R_{xy}=\frac{{\mathcal{T}_{II}-\mathcal{T}_{VI}}}{Q_{x}}$
and $R_{xx}=\frac{{\mathcal{T}_{II}-\mathcal{T}_{III}}}{Q_{x}}$ respectively.
As only the top surface is gapless, the heat current can only occur
on the top surface. The thermal resistivities satisfy $\rho_{xy}=R_{xy}$
and $\rho_{xx}=\frac{R_{xx}}{L_{x}/L_{y}}$. The thermal Hall conductance
of the system can be obtained through $\kappa_{xy}=\frac{\rho_{xy}}{\rho_{xy}^{2}+\rho_{xx}^{2}}$.

Fig. \ref{fig:3}(c) illustrates the variation of Hall thermal conductance
$\kappa_{xy}$ with increasing dephasing strength $\Gamma_{v}$.
At sufficiently
large dephasing strength, the Hall thermal conductance converges
to 1/4, with units the quantum thermal conductance $\kappa_0$.
There are two main reasons to consider the dephasing effect here.
Firstly, in real experiments, the size of samples often reach the
order of several hundred micrometers, far surpassing the decoherence
length, it is necessary to take the dephasing effect into consideration \citep{mogi2022experimental}.
Secondly, the chiral transport channels can manifest when the
dephasing strength is large enough \citep{zhou2022transport}.

Next,
we provide an explanation for the origin of the quarter-quantized
thermal Hall effect under dephasing.
The thermal Hall conductance is closely connected to
the chiral transport channels.
Chiral channels break spatial inversion symmetry.
For instance, considering the transmission probability
between site $a$ to site $b$, if there is a chiral channel between them,
it implies $T_{ab} \not= T_{ba}$, where $T_{ab}$ ($T_{ba}$)
represents the transmission probability from $a$ to $b$ ($b$ to $a$).
In contrast to  chiral p-wave superconductor \cite{sato2017topological}
where only 1D chiral Majorana edge mode exist within the energy gap,
here the gapless Majorana surface state on top surface
serves as a 2D thermal conductor.
This involves contributions from both isotropic 2D bulk (top surface) transport and 1D boundary chiral transport.
The calculation of the difference in transmission probabilities cancel
the isotropic contributions, allowing the extraction of boundary chiral transport.
To observe the spatial distribution of chiral channels,
the system should be partitioned into sufficiently large blocks,
ensuring that decoherence exists between these blocks due to their size being much larger than the coherence length.
The calculation of the difference in transmission probabilities
between blocks provides insight into the spatial distribution of chiral transport channels.
As the transmission probability
$T_{nm}$ between lead $n$ and lead $m$ decreases sharply with the
increasing distance $r_{nm}$ (see Fig.\ref{fig:3-1} in Appendix \ref{network}), the critical length can be defined
as $T_{nm}=0.001T_{neigb}$ \citep{xing2008influence}, where $T_{neigb}$
is defined as the sum of the transmission coefficient of electrons
and the cross Andreev process between the nearest neighboring leads.
We can use black boxes
to demarcate these regions of size $r_{c}\times r_{c}$ [Fig. \ref{fig:3}(b)].
Due to $r_{c}$ being much larger than the phase coherent length \citep{xing2008influence},
the transmission between two adjacent black boxes
becomes incoherent.
We can define the difference of transmission coefficients
between two black boxes, denoted as $T_{d}(x,y)=T_{ba}-T_{ab}$, where
$(x,y)$ represents the spatial coordinates of the lower right corner
site in box $a$. Here, $T_{ab}=\sum_{n\in a,m\in b}T_{nm}$ represents
the transmission probability from box $a$ to box $b$.
Fig. \ref{fig:3}(d) illustrates the spatial distribution of $T_{d}$,
revealing a transmission probability difference along the upper
and lower boundaries: $T_{d}=\pm1/4=\pm t_{\text{chiral}}$.
This implies the existence of quarter-quantized heat chiral channels
along the upper and lower boundaries, serving as the origin of the
quarter-quantized thermal Hall effect in Fig. \ref{fig:3}(c).

To further illustrate the connection between the quarter-quantized
chiral channels and the quarter-quantized thermal Hall effect here,
we partition the top surface of the system (where single massless Majorana
fermion exists) into blocks of size $r_{c}\times r_{c}$ [black box
in Fig.\ref{fig:3}(b)]. Due to the decoherence between these blocks,
the system can be effectively represented as a conductor-network model
that describes the transport in classical metals. Through analytical
solutions of the network model (see Appendix \ref{network}),
we establish the relationship
between the thermal Hall conductance $\kappa_{xy}$ and the difference
in transmission probability at the edge $t_{\text{chiral}}$:
\[
\kappa_{xy}= \kappa_0 \left[1+\frac{r_{c}\alpha t_{\text{chiral}}+t_{n}^{2}}{t_{\text{chiral}}^{2}+t_{n}^{2}}\frac{1}{L_{y}}
\right] t_{\text{chiral}}
\]
Here, $t_{n}=\frac{t_{13}+t_{12}+t_{14}}{r_{c}}$,
$\alpha=t_{56}+t_{57}+t_{89}+t_{810}-2t_{n}$, where
$t_{ij}=\sum_{n\in i,m\in j}T_{nm}x_{nm}$ [see Fig.\ref{fig:3}(b)], $x_{nm}=x_{n}-x_{m}$
and $x_{n(m)}$ represent the x-coordinate of lead $n(m)$. The width
of the system $L_{y}$ correlates with the quarter-quantized thermal
Hall conductance. As $L_{y}$ approaches infinity, $\kappa_{xy}=t_{\text{chiral}}\kappa_{0}$.

\begin{figure*}
    \begin{centering}
    \includegraphics[scale=0.5]{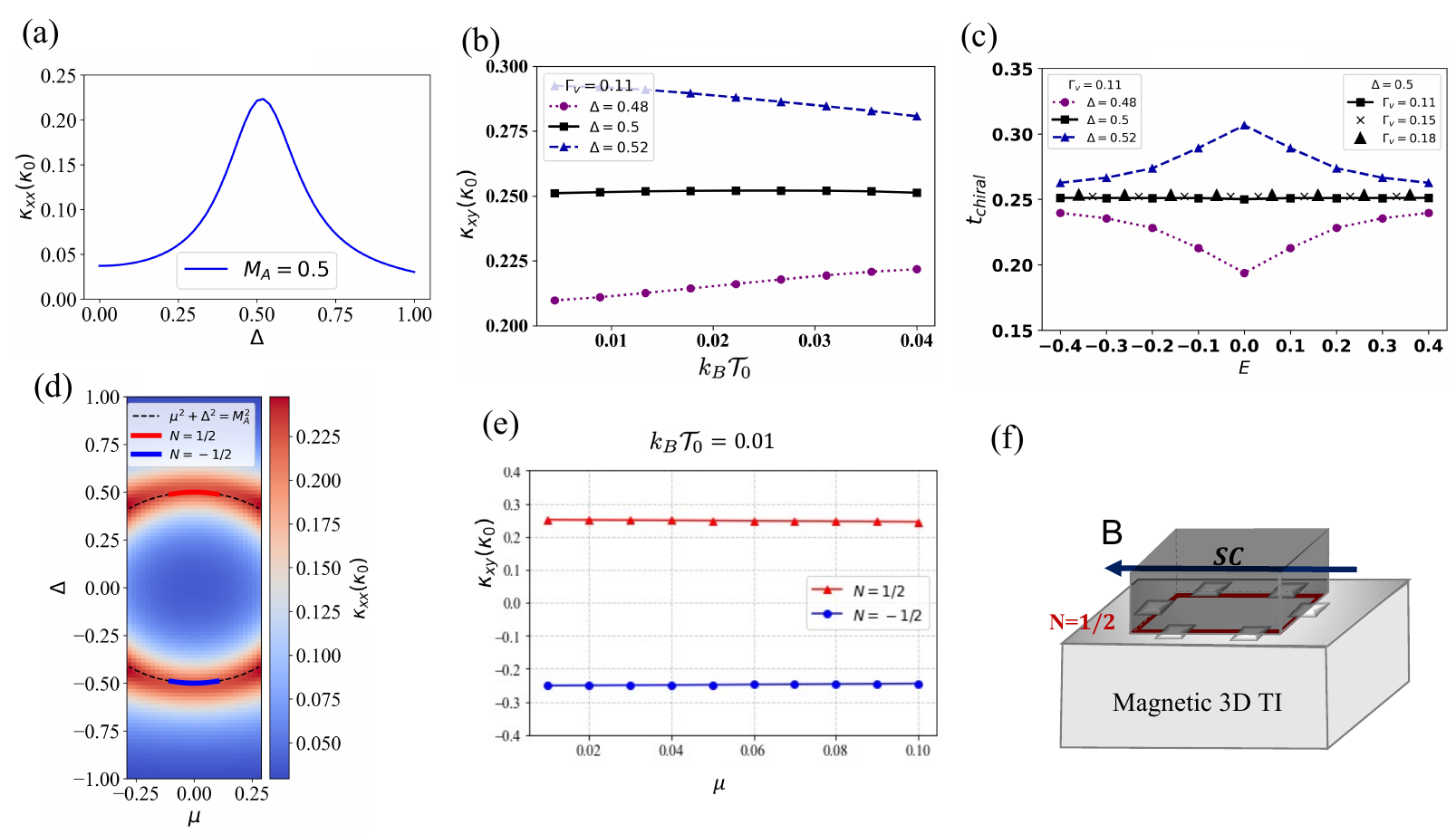}
    \par\end{centering}
    \caption{\label{fig:4}(a) For $\mu=0$, the longitudinal thermal conductance
    $\text{\ensuremath{\kappa_{xx}}}$ varies with the surface superconducting pairing potential $\Delta$.
    (b) The Hall thermal conductance changes
    with the background temperature $k_B \mathcal{T}_0$ for different $\Delta$.
    (c) { The transmission probability difference $t_{\text{chiral}}$
    varies with energy $E$ for different $\Delta$ and different dephasing strength $\Gamma_v$}.
    (d) Longitudinal thermal conductance $\kappa_{xx}$
    varying with $\Delta$ and $\mu$, with the dashed line representing
    the theoretically predicted phase transition curve.
    (e) Thermal Hall conductance $\kappa_{xy}$ on the red and blue curves in graph (d), with a temperature of $k_{B}\mathcal{T}_0=0.01$.
    (f) Experiment setup to realizing parity anomaly in gapless Majorana surface state through a magnetic 3D TI-superconductor heterostructure. The superconductor
    (deep gray) is overlaid on the surface of the magnetic 3D TI.
    By adjusting the superconducting pairing potential by a
    magnetic field, when gapless Majorana surface state appears, the boundary exhibits $N=1/2$ Majorana
    edge chiral states (represented by red lines).
    A quarter of the thermal Hall effect can be measured
    through a six-port Hall bar system (silver).
    { In the calculations, $L_{x}=50$, $L_{y}=50$,
    and $L_{z}=5$, with $\Gamma_v=0.15$ adopted in Fig. (a),(d) and (e), while $L_{x}=70$, $L_{y}=70$,
    and $L_{z}=5$ in (b) and (c).}}
    \end{figure*}

\section{\label{sec:temperature}Temperature dependence of thermal Hall conductance}

Our model relies on fine-tuning to align
the superconducting pairing potential $\text{\ensuremath{\Delta}}$
with the surface magnetization $M_{A}$, representing critical points
in the phase transition.
However, the phase transition point here
exhibits distinctive features that can be captured in experiments.
Since the realization of gapless Majorana surface state
is essentially a phase transition point
between $N=0$ and $N=1$ \citep{yan2021amajorana},
the states before and after are either insulating or 1D thermal conductors.
At the phase transition point, where the system transforms into a
2D thermal conductor, a distinctive peak emerges in the longitudinal
thermal conductance $\kappa_{xx}$. Fig. \ref{fig:4}(a) depicts the
variation of $\text{\ensuremath{\kappa_{xx}}}$ calculated using the
Landauer-B\"{u}ttiker formula with changing superconducting pairing
potential $\Delta$. A pronounced peak appears when $\Delta=M_{A}$.
In Fig. \ref{fig:4}(b), the temperature-dependent behavior of
the Hall thermal conductance $\kappa_{xy}$ for different $\Delta$
is presented (more details see Appendix \ref{temperature}).
Here, we consider $\ensuremath{{k_{B}}T\ll\Delta}$, allowing us to neglect
the influence of temperature on $\text{\ensuremath{\Delta}}$, which
can be solely controlled by the magnetic field. Notably, for the case
of $\Delta=M_{A}$, $\kappa_{xy}$ exhibits a quantized plateau under
temperature variations. However, as $\Delta$ deviates from $M_{A}$
[e.g., $\Delta=1.04M_{A}$ and $0.96M_{A}$, dashed and dotted lines
in Fig. \ref{fig:4}(b)],
not only does $\kappa_{xy}$ lose its quarter quantization,
but the plateau disappears as well.

The Hall thermal conductance at finite temperature depends on the all transport channels for energy scale of a few $k_B \mathcal{T}_0$ near the Fermi level.
This relationship is expressed as $\kappa_{xy}\text{\ensuremath{\left(\mathcal{T}_{0}\right)}}=\frac{1}{h}\int t_{chiral}(E)E\frac{\text{\ensuremath{\partial f_{0}}}}{\partial\mathcal{T}}dE$, where $f_{0}(E)=1/\left[e^{E/k_{B}\mathcal{T}_{0}}+1\right]$ is the Fermi
distribution (see Appendix \ref{temperature} for details). When $\text{\ensuremath{\Delta=M_{A}}}$, a single gapless Majorana cone exist on the surface. In this case, the chiral transmission coefficients, $t_{\text{chiral}}(E)$, which directly determines the thermal Hall conductance $\kappa_{xy}$,
remains energy-independent and is equal to 1/4 [Fig. \ref{fig:4}(c)]. 
Consequently, the 1/4 thermal Hall plateau persists under temperature variations [Fig. \ref{fig:4}(b)], providing evidence for the existence of a single gapless Majorana cone. { It's important to note that the role of dephasing here is to reduce the coherence length of the system. Even if the dephasing strength varies, as long as the coherence length of the system is much smaller than the system width $L_y$, $t_{\text{chiral}}$ remains energy-independent and stays close to 1/4 [See Fig. \ref{fig:4}(c)].}
When $\Delta$ deviates from $M_{A}$ (e.g., $\Delta=1.04M_{A}$ and $0.96M_{A}$), $t_{\text{chiral}}(E)$ gradually becomes energy-dependent [Fig. \ref{fig:4}(c)],
leading to $\kappa_{xy}$ depends on the temperature $\mathcal{T}_0$ 
also [Fig. \ref{fig:4}(b)].

The preceding analysis is based on the result with chemical potential
$\mu=0$, representing a specific case where electrons precisely fill
the Dirac point. While experimental means exist to adjust $\mu$,
achieving the exact filling of electrons at the Dirac point remains
challenging.
To better align with experimental conditions, we consider
the influence of $\mu$. In the case of a non-zero chemical potential,
the phase transition point satisfies $\ensuremath{{M_{A}}^{2}={\Delta^{2}}+{\mu^{2}}}$.
Figure \ref{fig:4}(d) illustrates the longitudinal thermal conductance
$\kappa_{xx}$ varying with $\Delta$ and $\mu$ at the fixed $M_A$.
The presence of 2D thermal conductor at the phase transition point
results in a peak in $\kappa_{xx}$.
We extract two intervals along the phase transition
curve to compute the corresponding Hall thermal conductance $\kappa_{xy}$ [indicated by red and blue dashed lines in Fig.\ref{fig:4}(d)].
$\kappa_{xy}$ maintains a quarter quantization with variation of
$\mu$, as shown in Fig.\ref{fig:4}(e).
Theoretically, the chemical potential $\mu$ induces coupling between
the two Majorana cones (see Appendix \ref{mu} for details).
However, when $\ensuremath{\mu\ll{M_{A}}}$,
the gapless Majorana surface states on the top surface
can be well-preserved, so the quarter-quantized Hall thermal conductance
well survives still.

\section{\label{sec:discuss}Discussion and conclusion}

Realizing parity anomaly in Majorana system relies
crucially on the interplay between
the topological surface state and superconductivity as well as magnetism.
There are two approaches to achieving the coexistence of 2D topological
surface states, superconductivity, and magnetism. One relies on the
magnetic TI with surface proximity-induced superconductivity, naturally
hosting topological surface states with magnetism. Similar experimental
configurations have been experimentally realized very recently \citep{yi2023interfaceinduced,xu2022proximityinduced,chen2022superconducting}.
The experimental setup involves inducing s-wave superconductivity
on the surface of a magnetic 3D TI, and arranging six-terminal Hall
bar measurements along the superconducting boundary to study the thermal
transport properties of the surface [see Fig.\ref{fig:4}(f)].
By adjusting the magnetic field to tune the superconducting gap ($\Delta$),
a peak in $\kappa_{xx}$ will be observed at the phase transition
point.
When chemistry potential $\mu$ near the Dirac point, the gapless Majorana surface state emerges
at the phase transition point, accompanied by a temperature-dependent
quarter-quantized Hall thermal platform at low temperatures.
Another approach is topological iron-based superconductors.
Coexistence of superconductivity and
topological surface states is observed in materials
like $FeTe_{0.55}Se_{0.45}$ \citep{zhang2018observation,zhang2019multiple},
with experiments confirming 0D Majorana states on the vortex of the
surface \citep{wang2018evidence,kong2019halfinteger,zhu2020nearlyquantized,liu2020anew}.
Here, we propose that by coupling a magnetic atomic layer to the surface,
providing surface magnetism, 2D Majorana surface states can be realized
in such materials. The Fermi level of $FeTe_{0.55}Se_{0.45}$ is very
close to the Dirac point \citep{wang2018evidence}. Simultaneously,
the superconducting gap on the surface is 1.8 meV \citep{wang2018evidence},
a magnitude comparable to the magnetic gap provided by magnetic proximity
effects \citep{bhattacharyya2021recentprogress}, which enhances the
feasibility of the experiment.

In summary, we present an approach based on 2D topological surface
state to achieve 2D Majorana surface states.
We analytically and numerically
demonstrate that a system with a single massless Majorana possesses
an $N=1/2$ Majorana edge current and a quarter-quantized Hall thermal
conductance. This signifies the presence of a parity anomaly in the
system, resembling the superconducting version of quantum anomalous
semimetal. By mapping the system to a network model, we elucidate
that the appearance of the 1/4 thermal Hall plateau under dephaseing
effect originates from the 1/4 chiral channels at the boundary. Furthermore,
we verify $\kappa_{xy}$ robustly quarter-quantized within a range
of background temperature $\mathcal{T}_{0}$. We also identify magnetic
topological insulator and topological iron-based superconductors $FeTe_{0.55}Se_{0.45}$
as potential platforms for realizing Majorana surface states.
Our study provides a framework
for the experimental realization of parity anomaly in Majorana system.\\

\section{\label{sec:Acknowledgment}Acknowledgment}
 Y.-H. W. is grateful to Hu-Mian
Zhou, Qing Yan, and Ludan Zhang for fruitful discussions.
This work was financially supported by the National Natural
Science Foundation of China (Grant No. 12374034 and No. 11921005),
the Innovation Program for Quantum Science and Technology
(2021ZD0302403), and the Strategic priority Research
Program of Chinese Academy of Sciences (Grant No.
XDB28000000). We also acknowledge the Highperformance
Computing Platform of Peking University for
providing computational resources.\\

\appendix

\section{\label{network}Landuer-buttiker formula and thermal Hall conductance}

\begin{figure}
\begin{centering}
\includegraphics[scale=0.5]{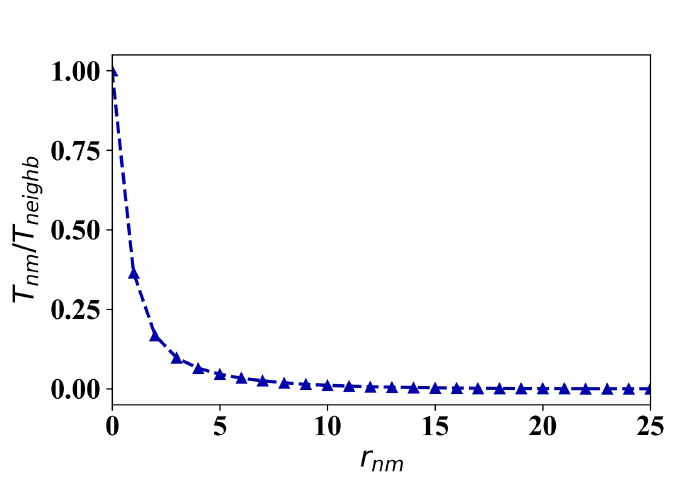}
\par\end{centering}
\caption{\label{fig:3-1} $T_{nm}/T_{neighb}$ vs. the distance $r_{nm}$ with $\Gamma_v=0.2$. }
\end{figure}

We consider a surface with dimensions $L_{x}\times L_{y}$ hosting
gapless Majorana surface state.
This surface is partitioned into a system of blocks, each with
dimensions $r_{c}\times r_{c}$ [see Fig.\ref{fig:3}(b)].
As the distance $r_{nm}>r_{c}$, the transmittance
probability $T_{nm}$ between lead $n$ and $m$ becomes nearly zero (see Fig.\ref{fig:3-1}),
resulting in non-zero transmittance only between neighboring blocks
[indicated by arrows in the Fig.\ref{fig:3}(b)].
We focus on the total heat flow
$Q_{x}$ in the horizontal direction within the central region of
the system.
\begin{widetext}
\begin{align}
Q_{x} & =\sum_{n\in left,m\in right}Q_{nm}=\sum_{n\in left,m\in right}\left(T_{nm}\mathcal{T}_{m}-T_{mn}\mathcal{T}_{n}\right)\kappa_0
\nonumber \\
 & =\sum_{n\in left,m\in right} T_{nm}\left(\mathcal{T}_{m}-\mathcal{T}_{n}\right)\kappa_0
 +\sum_{n\in left,m\in right}\left(T_{nm}-T_{mn}\right)\mathcal{T}_{n}
 \kappa_0 \nonumber \\
 & =\sum_{n\in left,m\in right}T_{nm}\left(x_{nm}\mathcal{\nabla T}_{x}+y_{nm}\mathcal{\nabla T}_{y}\right)\kappa_0 +\text{\ensuremath{\sum_{n\in6,m\in5}}}\left(T_{nm}-T_{mn}\right)
 \mathcal{T}_{n}\kappa_0  +\text{\ensuremath{\sum_{n\in9,m\in8}}}\left(T_{nm}-T_{mn}\right)
 \mathcal{T}_{n}\kappa_0\label{eq:Qx1}
\end{align}
\end{widetext}
Here, $x_{nm}=x_{n}-x_{m}$ and $y_{nm}=y_{n}-y_{m}$, where $x_{n(m)}$
and $y_{n(m)}$ represent the x-coordinate and y-coordinate of lead
$n(m)$. We assume that in the central region of the system, where
steady state is reached, the temperature gradient is constant.
$T_{nm}$ denotes the transmittance probability from lead $m$ to lead $n$,
with the assumption that $T_{nm}$ has translational symmetry except
at the boundaries.

The first term of Eq. (\ref{eq:Qx1}), $\sum_{n\in left,m\in right}T_{nm}\left(x_{nm}\mathcal{\nabla T}_{x}+y_{nm}\mathcal{\nabla T}_{y}\right)\kappa_0$, can
be categorized into two types: $L_{y}/r_{c}-2$ contributions from
the bulk [indicated by white dashed arrows in Fig.\ref{fig:3}(b)]
and two contributions
from the boundaries [indicated by red and blue arrows in Fig.\ref{fig:3}(b)].
The contribution from the bulk can be expressed as $((L_{y}/r_{c})-2)(t_{13}+t_{12}+t_{14})\nabla T_{x}$,
where $t_{ij}=\sum_{n\in i,m\in j}T_{nm}x_{nm}$ [see Fig.\ref{fig:3}(b)].
Note that terms related to $y_{nm}$ disappear due to ${T_{13}y_{13}+T_{14}y_{14}}=0$.
The contribution from the edges can be written as $(t_{56}+t_{57}+t_{89}+t_{810})\nabla T_{x}$.
Therefore, Eq. (\ref{eq:Qx1}) can be expressed as:
\begin{widetext}
\[
\begin{aligned}Q_{x} & =\left[(\frac{L_{y}}{r_{c}}-2)(t_{13}+t_{12}+t_{14})\mathcal{\nabla T}_{x}+(t_{56}+t_{57}+t_{89}+t_{810})\mathcal{\nabla T}_{x}+t_{chiral}\mathcal{T}_{up}-t_{chiral}\mathcal{T}_{low}
\right]\kappa_0\\
 & =[(L_{y}-2r_{c})t_{n}\mathcal{\nabla T}_{x}+r_{c}\alpha\mathcal{\nabla T}_{x}+t_{chiral}\mathcal{T}_{up}-t_{chiral}\mathcal{T}_{low}]
 \kappa_0\\
 & =[\left(t_{n}L_{y}+r_{c}\alpha\right)\mathcal{\nabla T}_{x}+t_{chiral}(\mathcal{T}_{up}-\mathcal{T}_{low})]
 \kappa_0\\
 & =[\left(t_{n}L_{y}+r_{c}\alpha\right)\mathcal{\nabla T}_{x}+t_{chiral}\mathcal{\nabla T}_{y}L_{y}]\kappa_0
\end{aligned}
\]
  \end{widetext}
The expression for $t_{n}$ is defined as $t_{n}=\frac{t_{13}+t_{12}+t_{14}}{r_{c}}$,
and$\alpha=t_{56}+t_{57}+t_{89}+t_{810}-2t_{n}$. Here, $\mathcal{T}_{\textup{up/down}}$
represents the temperatures at the upper and lower boundaries, satisfying
$\mathcal{T}_{\textup{up}}-\mathcal{T}_{\textup{low}}=\nabla{\mathcal{T}_{y}}\cdot L_{y}$. Similarly,
the heat flow in the $y$ (vertical) direction is given by
\[
Q_{y}=\left[\left(t_{n}L_{x}+r_{c}\alpha\right)\mathcal{\nabla T}_{y}-t_{\textup{chiral}}\mathcal{\nabla T}_{x}L_{x}\right]\kappa_0.
\]

Considering $Q_{y}=0$ and $L_{x}\gg0$, we obtain $t_{n}\mathcal{\nabla T}_{y}=t_{\textup{chiral}}\mathcal{\nabla T}_{x}$.
Due to $\rho_{xx}=L_{x}\mathcal{\nabla T}_{x}/Q_{x}$ and $\rho_{xy}=L_{y}\mathcal{\nabla T}_{y}/Q_{x}$,
the expression for $\kappa_{xy}$ can be derived as
\begin{align*}
\kappa_{xy} & =\frac{\rho_{xy}}{\rho_{xy}^{2}+\rho_{xx}^{2}}\\
 &
 =\left[1+\frac{r_{c}\alpha t_{chiral}+t_{n}^{2}}{t_{chiral}^{2}+t_{n}^{2}}\frac{1}{L_{y}}\right]
 t_{chiral}\kappa_0 .
\end{align*}

\section{\label{temperature}Hall thermal conductance with finite temperature}

In this Appendix, we will derive the Hall thermal conductance at finite
temperature. Starting from the non-zero temperature Landauer-Buttiker
formula:
\[
Q_{n}=\sum_{m}\widetilde{T}_{nm}(\mathcal{T}_{n}-\mathcal{T}_{m})
\kappa_0\]
where $\widetilde{T}_{nm}=\frac{3}{\pi^{2}k_{\mathrm{B}}^{2}\mathcal{T}_{0}}\int\left[T_{nm}(E)+T_{nm}^{\mathrm{CAR}}(E)\right]E\frac{\partial f_{0}}{\partial\mathcal{T}}dE$.
The finite temperature modifies the transmission probability
$\ensuremath{T_{nm}}(E)$ to $\widetilde{T}_{nm}$,
introducing contributions from different energy transmissions.
With the same mathematical structure,
following the method in Appendix \ref{network}, we obtain the total heat flows
in the horizontal (x) and vertical directions as: $Q_{x}=[\left(\widetilde{t}_{n}L_{y}+\widetilde{\alpha}\right)
\mathcal{\nabla T}_{x}+\widetilde{t}_{chiral}
\mathcal{\nabla  T}_{y}L_{y}] \kappa_0$,
$\ensuremath{{Q_{y}}=[\left(\widetilde{t}_{n}L_{x}
+\widetilde{\alpha}\right)\mathcal{\nabla T}_{y}-\widetilde{t}_{chiral}\mathcal{\nabla T}_{x}L_{x}]\kappa_0}$.
Here, $\widetilde{t}_{n}(\mathcal{T}_{0})
=\frac{3}{\pi^{2}k_{\mathrm{B}}^{2}\mathcal{T}_{0}}\int
t_{n}(E)E\frac{\partial f_{0}}{\partial\mathcal{T}}dE$,
$\widetilde{t}_{chiral}(\mathcal{T}_{0})
=\frac{3}{\pi^{2}k_{\mathrm{B}}^{2}\mathcal{T}_{0}}\int t_{chiral}(E)E\frac{\partial f_{0}}{\partial\mathcal{T}}dE$,
and $\widetilde{\alpha}=\frac{3}{\pi^{2}k_{\mathrm{B}}^{2}\mathcal{T}_{0}}
\int\alpha(E)r_{c}E\frac{\partial f_{0}}{\partial\mathcal{T}}dE$.
The Hall thermal conductance can be written as:
\[
\kappa_{xy}\left(\mathcal{T}_{0}\right)
=\kappa_0 \left[1+\frac{\widetilde{\alpha}\widetilde{t}_{chiral}
+\widetilde{t}_{n}^{2}}{\widetilde{t}_{chiral}^{2}+\widetilde{t}_{n}^{2}}
\frac{1}{L_{y}}\right]\widetilde{t}_{chiral}
\]
Considering $L_{y}\gg0$, the expression simplifies to:
\[
\kappa_{xy}\text{\ensuremath{\left(\mathcal{T}_{0}\right)}}=
\widetilde{t}_{chiral}\kappa_0
=\frac{1}{h}\int t_{chiral}(E)E\frac{\text{\ensuremath{\partial f_{0}}}}{\partial\mathcal{T}}dE
\]

\section{\label{mu}Majorana surface state with the finite chemistry potential}

When considering the chemical potential $\mu$, the critical point
satisfies $M_{A}^{2}=\Delta^{2}+\mu^{2}$ \cite{yan2021amajorana}.
The Hamiltonian with Majorana
basis $\Psi_{\mathbf{k}}=\ensuremath{\frac{1}{\sqrt{2}}\left(c_{\mathbf{k}\uparrow}+c_{-\mathbf{k}\downarrow}^{\dagger},c_{\mathbf{k}\downarrow}+c_{-\mathbf{k}\uparrow}^{\dagger},-c_{\mathbf{k}\downarrow}+c_{-\mathbf{k}\uparrow}^{\dagger},-c_{\mathbf{k}\uparrow}+c_{-\mathbf{k}\downarrow}^{\dagger}\right)^{T}}$
can be written as:
\begin{align*}
H & =\text{\ensuremath{\left( \begin{array}{cccc}
\text{\ensuremath{\Delta}}+M_{A} & k_{x}-ik_{y} & 0 & -\text{\ensuremath{\mu}}\\
k_{x}+i\text{\ensuremath{k_{y}}} & -\text{\ensuremath{\Delta}}-M_{A} & -\text{\ensuremath{\mu}} & 0\\
0 & -\text{\ensuremath{\mu}} & \text{\ensuremath{\Delta}}-M_{A} & \text{\ensuremath{k_{x}}}+ik_{y}\\
-\mu\  & 0 & k_{x}-ik_{y} & M_{A}-\text{\ensuremath{\Delta}}
\end{array} \right)}}\\
 & =\text{\ensuremath{\left({\begin{array}{cc}
{H_{+}^ {}\left(k\right)} & {-\mu\sigma_{x}}\\
{-\mu\sigma_{x}^ {}} & {H_{-}^ {}\left(k\right)}
\end{array}}\right)}}
\end{align*}
The chemical potential $\mu$ induces coupling between the two Majorana
cones and simultaneously opens gaps in both cones, with gap sizes
of $2|\sqrt{M_{A}^{2}-\mu^{2}}+M_{A}|$ for $H_{+}(k)$ and $2|\sqrt{M_{A}^{2}-\mu^{2}}-M_{A}|$
for $H_{-}(k)$.
Next, we demonstrate that the Majorana surface state exists
when $\mu\ll M_{A}$.
At $\mu\ll M_{A}$, $H_{-}(k)$ can be written as:
\begin{align*}
H_{-}(k) & =k_{x}\sigma_{x}-k_{y}\sigma_{y}+(\sqrt{M_{A}^{2}-\mu^{2}}-M_{A})\sigma_{z}\\
 & \approx k_{x}\sigma_{x}-k_{y}\sigma_{y}
 -\frac{M_{A}}{2}(\frac{\mu}{M_{A}})^{2}\sigma_{z}
\end{align*}
When $\mu\ll M_{A}$, the gap can be approximated as a second-order
small quantity with respect to $\frac{\mu}{M_{A}}$.
Simultaneously, $\mu$ introduces coupling
between $H_{+}$ and $H_{-}$, and the
mass term of $H_{+}$ affects $H_{-}$ through this coupling.
Let $M_{+}=\sqrt{M_{A}^{2}-\mu^{2}}+M_{A}$. Considering the mass term
of $H_{+}$ and its self-energy correction to $H_{-}$, the expression
is as follows:
\begin{align*}
\Sigma(E)&=-\mu\sigma_{x}\frac{1}{E-M_{+}\sigma_{z}}(-\mu\sigma_{x})\\&=\frac{\mu^{2}M_{+}}{-E^{2}+M_{+}^{2}}\sigma_{z}-\frac{\mu^{2}E}{-E^{2}+M_{+}^{2}}\sigma_{0}
\end{align*}
Since $M_{+}>M_{A}$, $\frac{\mu}{M_{+}}$ is also a small quantity.
Therefore, the mass term correction introduced by $H_{+}$ to $H_{-}$
is also a second-order small quantity. In summary, when $\mu\ll M_{A}$,
the gapless Majorana Dirac cone can exist.
In experiments, where $M_{A}$ is typically $10$ meV  \cite{schmidt2011spintexture,henk2012topological},
making it possible to manipulate $\mu$ within the range of $0.1M_{A}$
by tuning the chemical potential.

\bibliography{ref}

\end{document}